\documentclass[preprint,12pt]{elsarticle}
 \usepackage{graphicx}
\usepackage{amssymb}
\usepackage{amsthm}

\journal{Physica A}

\begin{document}

\begin{frontmatter}

\title{
Fisher-information and the thermodynamics of scale-invariant systems}

\author[a1]{A. Hernando}
\ead{alberto@ecm.ub.es}
\author[a2]{C. Vesperinas}
\ead{cristina.vesperinas@sogeti.com}
\author[a3]{A. Plastino}
\ead{plastino@fisica.unlp.edu.ar}
\address[a1]{Departament ECM, Facultat de F\'{\i}sica,
Universitat de Barcelona. Diagonal 647, 08028 Barcelona, Spain}
\address[a2]{Sogeti Espa\~na, WTCAP 2, Pla\c ca de la Pau s/n, 08940
Cornell\`a, Spain}
\address[a3]{National University La Plata, IFCP-CCT-CONICET, c.c. 727, 1900 La Plata, Argentina}

\begin{abstract}

We present a thermodynamic formulation for scale-invariant systems
based on the minimization with constraints of Fisher's information
measure. In such a way a clear analogy between these systems's
thermal properties and those of  gases and fluids is seen to emerge
in natural fashion. We focus attention on the non-interacting
scenario, speaking thus  of \emph{scale-free ideal gases} (SFIGs)
and present some empirical evidences regarding such disparate
systems as electoral results, city populations and total citations
in Physics journals,  that seem to indicate that SFIGs do exist. We
also illustrate the way in which Zipf's law can be understood in a
thermodynamical context as the surface of a finite system. Finally,
we derive an equivalent microscopic description of our systems which
totally agrees with previous numerical simulations found in the
literature.

\end{abstract}

\begin{keyword}
Fisher information \sep scale-invariance
\end{keyword}

\end{frontmatter}

\section{Introduction}

Scale-invariant phenomena are rather abundant in Nature and display
somewhat unexpected features. Examples can be found that range from
physical and biological to technological and social
sciences~\cite{uno}. One may cite, among many possibilities, that
empirical data from percolation theory and nuclear
multifragmentation~\cite{perco} reflect scale-invariant behaviour,
as does the abundance of genes in various organisms and
tissues~\cite{furu}. Additionally, we can speak of   the frequency
of words in natural languages~\cite{zip}, scientific collaboration
networks~\cite{cites}, the Internet traffic~\cite{net1}, Linux
packages links~\cite{linux}, as well as of electoral
results~\cite{elec1,ccg}, urban agglomerations~\cite{ciudad,ciudad2} and
firm sizes all over the world~\cite{firms}. What characterizes these
disparate systems  is the lack of a characteristic size, length or
frequency for an observable $k$ under scrutiny. This fact usually
leads to a power law distribution $p(k)$, valid in most of the
domain of definition of $k$,
\begin{equation}\label{eq1}
p(k)\sim1/k^{1+\gamma},
\end{equation}
with $\gamma\geq0$. Special attention deserves the class of
universality defined by $\gamma=1$, which corresponds to the
so-called  Zipf's law in the cumulative distribution or the
rank-size
distribution~\cite{perco,furu,zip,net1,linux,ciudad,ciudad2,firms,citis}.
Recently, Maillart et al.~\cite{linux} have found that  links'
distributions follow Zipf's law as a consequence of stochastic
proportional growth. In its simplest formulation such kind of growth
assumes that an element of the system becomes enlarged
proportionally to its size $k$, being governed by a Wiener process.
The class $\gamma=1$ emerges from the condition of stationarity,
i.e., when the system reaches a dynamic equilibrium~\cite{citis}.
We will as well propose to consider the case $\gamma=0$ as representative of a second class of universality,
since the ensuing behavior, empirically found by Costa Filho et al.~\cite{elec1} with regards to
the vote-distribution in  Brazilian electoral results, emerges 
as the result of multiplicative processes in complex networks \cite{ccg}.

In this paper we attempt to formulate a thermodynamic treatment
common to these systems. Our efforts are based on the minimization
with appropriate constraints of Fisher's information measure (FIM),
abbreviated as the MFI approach. It is shown in \cite{fisher2} that
MFI leads to a (real) Schreodinger-like equation whose ``potential''
function is given by the constraints employed to constrain the
variational process. The interplay between constraints and
associated Lagrange multipliers turns our to be Legendre-invariant
\cite{fisher2} and leads to all known thermodynamic relations. Such
result constitutes the essential ingredient of our considerations
here. We will first consider the MFI treatment of the ideal gas (Seccion 3), not given elsewhere as far as we are aware of, 
since it is indispensable to deal with it in order to 
fully understand the methodology employed for scale-free systems, which is tackled in Section 4. Applicatioms are discussed in Section 5 and some conclusiones drawn in Section 6. We begin our considerations in Section 2 with a brief Fisher's sketch.

\section{Minimum Fisher Information approach (MFI)}
\label{p2}

The Fisher information measure $I$ for a system described by a set
of coordinates $\mathbf{q}$ and physical parameters
$\mathbf{\theta}$, has the form~\cite{libro}
\begin{equation}\label{fish}
I(F)=\int_\Omega
d\mathbf{q}F(\mathbf{q}|\mathbf{\theta})\sum_{ij}c_{ij}\frac{\partial}{\partial\theta_i}\ln
F(\mathbf{q}|\mathbf{\theta})\frac{\partial}{\partial\theta_j}\ln
F(\mathbf{q}|\mathbf{\theta}),
\end{equation}
where $F(\mathbf{q}|\mathbf{\theta})$ is the density distribution in
a configuration space ($\mathbf{q}$) of volume $\Omega$ conditioned
by the physical parameters collectively represented by the variable
$\mathbf{\theta}$. The constants $c_{ij}$ account for
dimensionality, and take the form $c_{ij}=c_i\delta_{ij}$ if $q_i$
and $q_j$ are uncorrelated, where
$\delta_{ij}$ is the Kronecker delta. As shown in~\cite{fisher2}, the
thermal-equilibrium state of the system can be determined by
minimizing $I$ subject to adequate prior conditions (MFI), like the
normalization of $F$ or by any constraint on the mean value of an
observable $\langle A_i \rangle$~\cite{fisher2}. The MFI is then
cast as a variation problem of the form
\begin{equation}
\delta\left\{I(F)-\sum_i\mu_i\langle A_i \rangle\right\}=0,
\end{equation}
where $\mu_i$ are appropriate Lagrange multipliers.

\section{MFI treatment of the ideal gas}
\label{aa}

As a didactic introductory example, not discussed in \cite{fisher2},
we will here rederive, {\it via MFI} (something original as far as
we know), the density distribution, in configuration space, of the
(translational invariant) ideal gas (IG)~\cite{termo}, that
describes non-interacting classical particles of mass $m$ with
coordinates $\mathbf{q}=(\mathbf{r},\mathbf{p})$, where
$md\mathbf{r}/dt=\mathbf{p}$. The translational invariance is
described by the translational family of distributions
$F(\mathbf{r},\mathbf{p}|\mathbf{\theta}_r,\mathbf{\theta}_p)=F(\mathbf{r}',\mathbf{p}')$
whose form does not change under the transformations
$\mathbf{r}'=\mathbf{r}-\mathbf{\theta}_r$ and
$\mathbf{p}'=\mathbf{p}-\mathbf{\theta}_p$. We assume that these
coordinates are canonical~\cite{mec} and uncorrelated. This
assumption is introduced into the information measure~(\ref{fish})
setting $c_{ij}=c_i\delta_{ij}$, where $c_{i}=c_r$ for space
coordinates, $c_{i}=c_p$ for momentum coordinates. The density can obviously be
factorized  in the fashion
$F(\mathbf{r},\mathbf{p})=\rho(\mathbf{r})\eta(\mathbf{p})$, and
then~\cite{libro} it follows from the additivity of the information
measure that $I=I_r+I_p$.  If $D$ is the dimensionality we have
\begin{equation}
\begin{array}{rl}
I_r=&\displaystyle c_r\int
d^D\mathbf{r}~\rho(\mathbf{r})\left|\mathbf{\nabla}_r \ln \rho(\mathbf{r})\right|^2\\
I_p=&\displaystyle c_p\int
d^D\mathbf{p}~\eta(\mathbf{p})\left|\mathbf{\nabla}_p\ln
\eta(\mathbf{p})\right|^2.
\end{array}
\end{equation}

In extremizing FIM we constrain the normalization of
$\rho(\mathbf{r})$ and $\eta(\mathbf{p})$ to the total number of
particles $N$ and to $1$, respectively, i.e.,
\begin{equation}\label{normg}
\int d^D\mathbf{r}~\rho(\mathbf{r})=N,\qquad\int
d^D\mathbf{p}~\eta(\mathbf{p})=1.
\end{equation}
In addition, we penalize infinite values for the particle momentum
with a constraint on the variance of $\eta(\mathbf{p})$ to a given
empirically obtained value, namely,
\begin{equation}\label{sp2}
\int
d^D\mathbf{p}~\eta(\mathbf{p})(\mathbf{p}-\overline{\mathbf{p}})^2=D\sigma_p^2,
\end{equation}
where $\overline{\mathbf{p}}$ is the mean value of $\mathbf{p}$. For
each degree of freedom it is known from the Virial Theorem that the
variance is related to the temperature $T$ as $\sigma_p^2=mk_BT$,
with $k_B$ the Boltzmann constant. Variation thus yields
\begin{equation}\label{exg}
\displaystyle\delta\left\{c_r\int
d^D\mathbf{r}~\rho\left|\mathbf{\nabla}_r \ln \rho\right|^2+\mu\int
d^D\mathbf{r}~\rho\right\}=0
\end{equation}
and
\begin{equation}\label{exh}
\delta\left\{c_p\int d^D\mathbf{p}~\eta\left|\mathbf{\nabla}_p\ln
\eta\right|^2+\lambda\int
d^D\mathbf{p}~\eta(\mathbf{p}-\overline{\mathbf{p}})^2 +\nu\int
d^D\mathbf{p}~\eta\right\}=0,
\end{equation}
where $\mu$, $\lambda$ and $\nu$ are Lagrange multipliers.
Introducing now $\rho(\mathbf{r})=\Psi^2(\mathbf{r})$ and
varying~(\ref{exg}) with respect to $\Psi$ leads to a
Schroedinger-like equation~\cite{fisher2,QM}
\begin{equation}
\left[-4\nabla_r^2+\mu'\right]\Psi(\mathbf{r})=0,
\end{equation}
where $\mu'=\mu/c_r$. To fix the boundary conditions, we first
assume that the $N$ particles are confined in a box of volume $V$,
and next we take the thermodynamic limit $N,V\rightarrow\infty$ with
$N/V$ finite. The equilibrium state compatible with this limit
corresponds to the ground state solution ($\mu'=0$), which is the uniform
density $\rho(\mathbf{r})=N/V$.

Introducing $\eta(\mathbf{p})=\Phi^2(\mathbf{p})$ and
varying~(\ref{exh}) with respect to $\Phi$ leads to the quantum
harmonic oscillator-like equation~\cite{QM}
\begin{equation}
\left[-4\nabla_p^2+\lambda'(\mathbf{p}-\overline{\mathbf{p}})^2+\nu'\right]\Phi(\mathbf{p})=0,
\end{equation}
where $\lambda'=\lambda/c_p$ and $\nu'=\nu/c_p$. The equilibrium
configuration corresponds to the ground state solution, which is now
a gaussian distribution. Using~(\ref{sp2}) to identify
$|\lambda'|^{-1/2}=\sigma_p^2$ we get the Maxwell-Boltzmann distribution,
which leads to a density distribution in configuration space of the
form
\begin{equation}
f(\mathbf{r},\mathbf{p})=\frac{N}{V}\frac{\exp\left[-(\mathbf{p}-\overline{\mathbf{p}})^2/2\sigma_p^2\right]}{(2\pi\sigma_p^2)^{D/2}}.
\end{equation}
If $H$ is the elementary volume in phase space, the total number of
microstates is
$Z=N!H^{DN}\prod_{i=1}^NF_1(\mathbf{r}_i,\mathbf{p}_i)$, where
$F_1=F/N$ is the monoparticular distribution and $N!$ counts all
possible permutations for distinguishable particles. The entropy
$S=-k_B\ln Z$ gets then written in the form
\begin{equation}\label{SIG}
S=Nk_B\left\{\ln\frac{V}{N}\left(\frac{2\pi
\sigma_p^2}{H^2}\right)^{D/2}+\frac{2+D}{2}\right\},
\end{equation}
where we have used the Stirling approximation for $N!$. This
expression agrees, of course with the known value entropic
expression for the IG~\cite{termo}, illustrating on the predictive
power of the MFI formulation advanced in \cite{fisher2}.\\

\section{Scale invariant systems}

We pass now to the leit-motif of the present communication and
consider a one-dimensional system with dynamical coordinates
$\mathbf{q}=(k,v)$ where $dk/d\tau=v$, with $\tau$ the time
variable. We define $k$ as a {\it discrete} coordinate, i.e.
$k=k_1,k_2,\ldots,k_M$, where $k_i=i\Delta k$ and $M\gg1$, is the
total number of bins of width $\Delta k$ in our system. In order to
address the scale-invariance behaviour of $k$ we change variables
passing to  new coordinates $u=\ln k$ and $w=du/dt$. We work under
the hypothesis that $u$ and $w$ are canonically
conjugated~\cite{mec} and uncorrelated. This assumption immediately
leads to proportional growth since
\begin{equation}\label{dyn}
d k/d t=v=kw.
\end{equation}
For constant $w$ this equation yields an exponential growth
$k=k_0e^{wt}$, which represents uniform linear motion in $u$, that
is, $u=wt+u_0$, with $u_0=\ln k_0$~\footnote{This exponential growth
allows to identify the systems that we study in this work in
macroscopic fashion with those addressed  in~\cite{exp}.}. It is
easy to check that the scale transformation $k'=k/\theta_k$ leaves
invariant the coordinate $w$, whereas the coordinate $u$ transforms
translationally as $u'=u-\Theta_k$, where $\Theta_k=\ln\theta_k$.
Thus, the physics does not depend on scale and the system is
translationally invariant with respect to the coordinates $u$ and
$w$, entailing that the distribution of physical elements can be
described by the monoparametric translation families
$f(u,w|\Theta_k,\Theta_w)=f(u',w')$. By analogy with the IG, we will
call our system a ``scale-free ideal gas'' (SFIG), i.e., a system of
$N$ non-interacting elements. Taking into account that i)
 $u$ and $w$ are canonical and uncorrelated ($c_{ii}=c_i\neq0$ and $c_{uw}=c_{wu}=0$), 
so the density distribution can be factorized as $f(u,w)=g(u)h(w)$,
and ii) that the Jacobian for our change of variables is
$dkdv=e^{2u}dudw$, the information measure $I=I_u+I_w$ can be
obtained in the continuous limit as
\begin{equation}
\begin{array}{rl}
I_u=&\displaystyle c_u\int_\Omega d u~e^{2u}g(u)\left|\frac{\partial\ln g(u)}{\partial u}\right|^2\\
I_w=&\displaystyle c_w\int_{-\infty}^\infty d
w~h(w)\left|\frac{\partial\ln h(w)}{\partial w}\right|^2,
\end{array}
\end{equation}
where $\Omega=\ln(k_M/k_1)=\ln M$ is the volume defined in ``$u$''-space.

\subsection{MFI treatment of the scale-free ideal gas} \label{p3}

The constraints to the given observables $\langle A_i \rangle$ in
the extremization problem determine the behaviour of the system. For
the general case, we constrain the normalization of $g(u)$ and
$h(w)$ to the total number of particles $N$ and to $1$, respectively
\begin{equation}\label{w2a}
\int_\Omega d u~e^{2u}g(u)=N,\qquad\int_{-\infty}^\infty d w~h(w)=1.
\end{equation}
In addition, we penalize infinite values for $w$ with a constraint
on the variance of $h(w)$ to a given measured value
\begin{equation}
\int_{-\infty}^\infty d
w~h(w)(w-\overline{w})^2=\sigma_w^2,\label{w2b}
\end{equation}
where $\overline{w}$ is the average growth. The variation yields
\begin{equation}\label{varg}
\delta\left\{c_u \int_\Omega d u~e^{2u}g\left|\frac{\partial\ln
g}{\partial u}\right|^2+\mu \int_\Omega d u~e^{2u}g\right\}=0
\end{equation}
and
\begin{equation}\label{varh}
\delta\left\{c_w \int_{-\infty}^\infty d w~h\left|\frac{\partial\ln
h}{\partial w}\right|^2+\lambda \int_{-\infty}^\infty d
w~h(w-\overline{w})^2 +\nu\int_{-\infty}^\infty d w~h\right\}=0,
\end{equation}
where $\mu$, $\lambda$ and $\nu$ are Lagrange multipliers.
Introducing $g(u)=e^{-2u}\Psi^2(u)$, and varying ~(\ref{varg}) with
respect to $\Psi$ leads, as is always the case with  the MFI
\cite{fisher2}, to the Schroedinger-like equation
\begin{equation}
\left[-4\frac{\partial^2}{\partial u^2}+4+\mu'\right]\Psi(u)=0,
\end{equation}
where $\mu'=\mu/c_u$. Analogously to the IG, we impose solutions
compatible with a finite normalization of $g$ in the thermodynamic
limit $N,\Omega\rightarrow\infty$ with $N/\Omega=\rho_0$ finite,
where $\rho_0$ is defined as the \emph{bulk density}. Solutions
compatible with the normalization of~(\ref{w2a}) are given by
$\Psi(u)=A_\alpha e^{-\alpha u/2}$, where $A_\alpha$ is the
normalization constant and $\alpha=\sqrt{4+\mu'}$. In this general
case, the density distribution as a function of $k$ takes the form
of a power law: $g_\alpha(\ln k)=A^2/k^{2+\alpha}$. The equilibrium
is always defined for the MFI as the ground state solution
\cite{fisher2}, which corresponds to the lowest allowed value
$\alpha=0$.

Introducing now $h(w)=\Phi^2(w)$ and varying~(\ref{varh}) with
respect to $\Phi$ leads to the quantum harmonic oscillator-like
equation~\cite{fisher2,QM}
\begin{equation}
\left[-4\frac{\partial^2}{\partial
w^2}+\lambda'(w-\overline{w})^2+\nu'\right]\Phi(w)=0,
\end{equation}
where $\lambda'=\lambda/c_w$ and $\nu'=\nu/c_w$. The equilibrium
configuration corresponds to the ground state solution, which is now
a Gaussian distribution. Using~(\ref{w2b}) to identify
$|\lambda'|^{-1/2}=\sigma_w^2$ we get the Maxwell-Boltzmann distribution
\begin{equation}\label{hw}
h(w)=\frac{\exp\left[-(w-\overline{w})^2/2\sigma_w^2\right]}{\sqrt{2\pi}\sigma_w}.
\end{equation}

The density distribution in configuration space
$F(k,v)dkdv=f(u,w)e^{2u}dudw$ is then
\begin{equation}
F(k,v)=\frac{N}{\Omega
k^2}\frac{\exp\left[-(v/k-\overline{w})^2/2\sigma_w^2\right]}{\sqrt{2\pi}\sigma_w}.\label{fkv}
\end{equation}
If we define $H=\Delta k^2/\Delta\tau$ as the elementary volume in
phase space, where $\Delta\tau$ is the time element, the total
number of microstates is $Z=N!H^N\prod_{i=1}^NF_1(k_i,v_i)$, where
$F_1=F/N$ is the monoparticular distribution function and $N!$
counts all possible permutations for distinguishable elements. The
entropy equation of state $S=-\kappa\ln Z$ reads
\begin{equation}
S=N\kappa\left\{\ln\frac{\Omega}{N}\frac{\sqrt{2\pi}\sigma_w}{H'}+\frac{3}{2}\right\},
\end{equation}
where $\kappa$ is a constant that accounts for dimensionality and
$H'=H/(k_M k_1)=H/(M\Delta k^2)=1/(M\Delta\tau)$. Remarkably, this
expression has the same form as the one-dimensional IG ($D=1$ in
~(\ref{SIG})); instead of the thermodynamical variables $(N,V,T)$,
here we deal with the variables $(N,\Omega,\sigma_w)$, which make
the entropy scale-invariant as well.

\begin{figure}[t!]
\centering
\includegraphics[width=0.5\linewidth,clip=true]{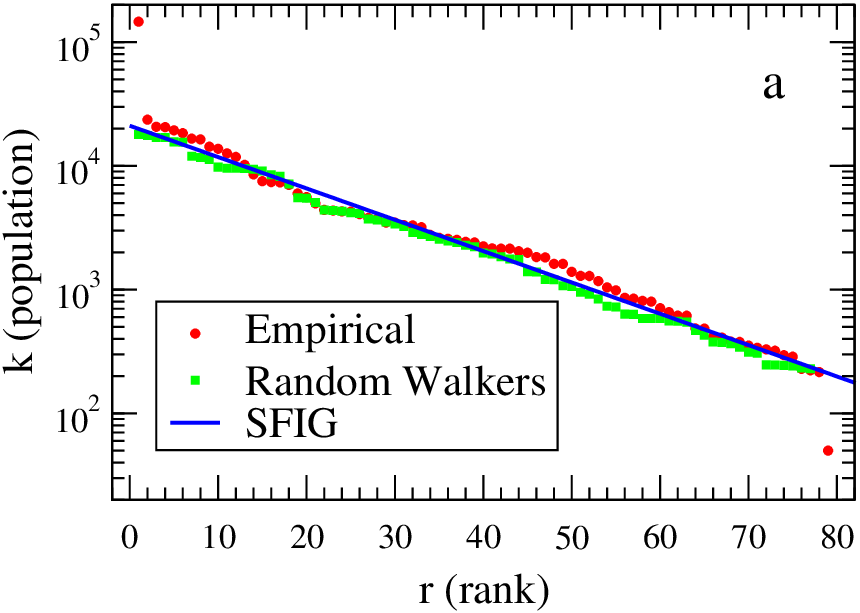}\\
\includegraphics[width=0.5\linewidth,clip=true]{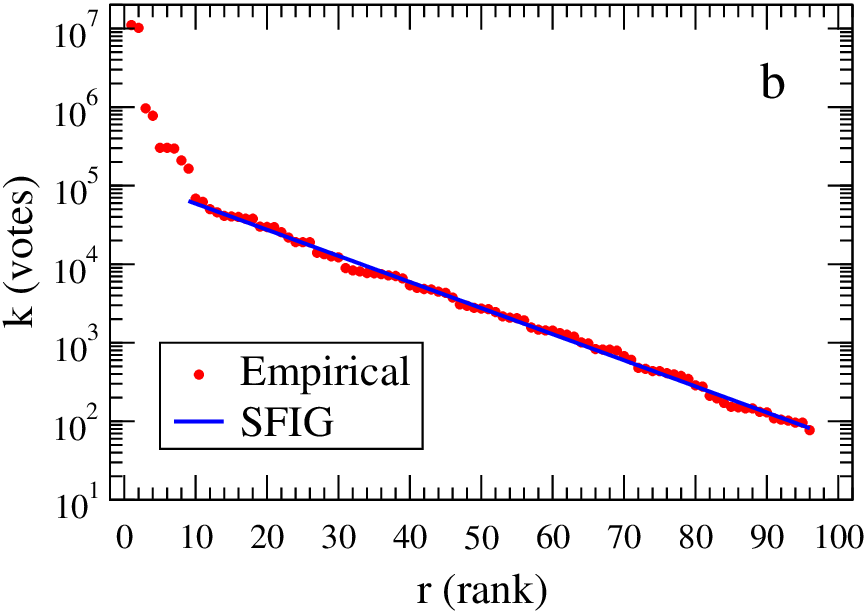}\\
\includegraphics[width=0.5\linewidth,clip=true]{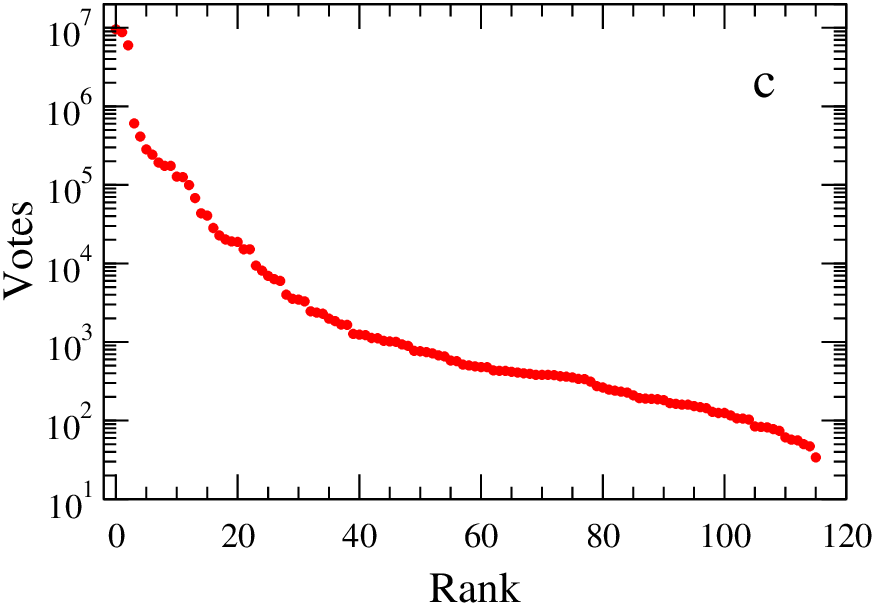}%
\caption{(colour on-line) \textbf{a}, rank-size distribution of the
cities of the province of Huelva, Spain (2008), sorted from largest
to smallest, compared with the result of a simulation with Brownian
walkers (green squares). \textbf{b}, rank-plot of the 2008 General
Elections results in Spain. \textbf{c}, rank-plot of the 2005
General Elections results in the United Kingdom. (Red dots:
empirical data; blue lines: fit to (\ref{coldr})).\label{fig1}}
\end{figure}

The total density distribution for $k$ is obtained integrating for
all $v$ the density distribution in configuration space.
Accordingly, from~(\ref{fkv}) we get
\begin{equation}\label{cold}
F(k)=\int dvF(k,v)=\frac{N}{\Omega}\frac{1}{k}=\frac{\rho_0}{k}.
\end{equation}
It can be shown  that this it is just a uniform density-distribution in 
$u$-space of the bulk density: $F(k)dk=f(u)e^udu=N/\Omega du=\rho_0du$.

\section{Social examples of scale-free ideal gases}

A common representation of empirical data is the so-called rank-plot
or Zipf plot~\cite{zip,ciudad2,mod1}, where the $j$th element of the
system is represented by its size, length or frequency $k_j$ against
its rank, sorted from the largest to the smallest one. This process
just renders the inverse function of the ensuing cumulative
distribution, normalized to the number of elements. We call $r$ the
rank that ranges from 1 to $N$. For large $N$, the density
distribution (\ref{cold}) correspond to an exponential rank-size
distribution
\begin{equation}\label{coldr}
k(r)=k_M\exp\left[-\frac{r-1}{\rho_0}\right].
\end{equation}

\begin{figure}[t!]
\centering
\includegraphics[width=0.5\linewidth,clip=true]{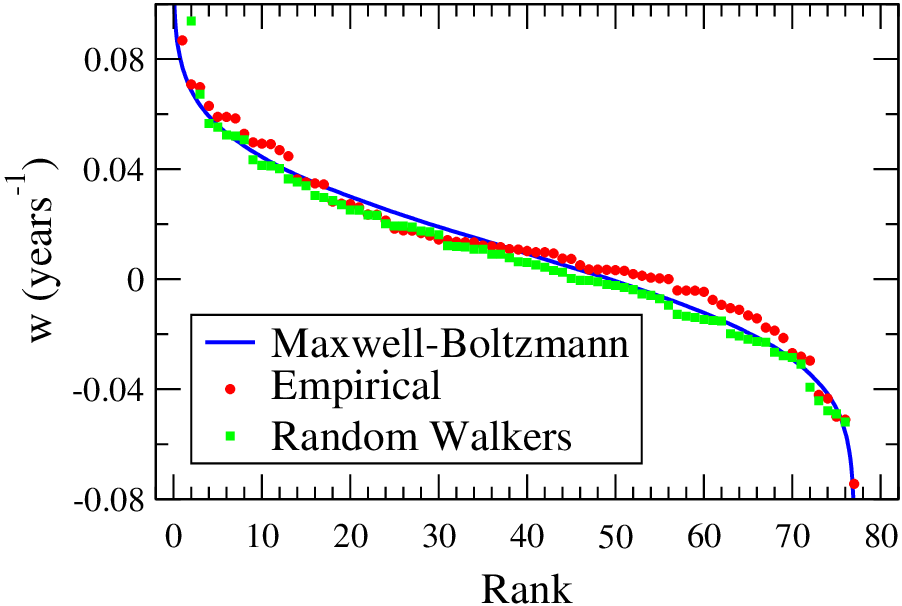}%
\caption{(colour on-line) Rank-plot of the growth rate $w$ of the
province of Huelva between 2007 and 2008 (red dots) compared with a
Boltzmann distribution with the same mean value and standard
deviation (blue line).\label{fig2}}
\end{figure}

This behaviour, which corresponds to the class of universality
$\gamma=0$ in~(\ref{eq1}), is that empirically found by Costa Filho
et al.~\cite{elec1} in the distribution of votes in the Brazilian
electoral results. We have found such a behaviour in i) the
city-size distribution of small regions (as in the province of
Huelva (Spain)~\cite{muni}) and ii)  electoral results (as in the
2008 Spanish General Elections results~\cite{elecSp}). We depict in
figures~1a and 1b the pertinent  rank-sizes distributions in
semi-logarithmic scale, where a straight line corresponds to a
distribution of type~(\ref{coldr}). A large portion of the
distributions can be fitted to (\ref{coldr}), with a correlation coefficient
of $0.994$ and $0.998$, respectively. From these fits we have obtain
a bulk density of $\rho_0=0.058$ for the General Elections results,
and in the case of Huelva of $\rho_0=17.1$ ($N=77$, $\Omega=4.5$).
Using historical data for the later~\cite{muni}, we have used the
backward differentiation formula to calculate the relative growth
rate of the $i$-th city as
\begin{equation}\label{wi}
w_i = \frac{\ln k_i^{(2008)}-\ln k_i^{(2007)}}{\Delta t}
\end{equation}
where $k_i^{(2007)}$ and $k_i^{(2008)}$ are the number of
inhabitants of the $i$-th city in $2007$ and $2008$, respectively
while $\Delta t=1$ year. We show in figure~\ref{fig2} the empirical
rank-plot of the relative growth, where we have obtained
$\overline{w}=0.011$~years$^{-1}$ and $\sigma_w=0.030$~years$^{-1}$,
compared with the rank-plot of a Maxwell-Boltzmann distribution with the
same mean value and standard deviation.

However, these regularities are not always obvious to the naked eye,
as shown for the case of the most voted parties in Spain'08 or for
the whole distribution of the 2005 General Elections results in the
United Kingdom~\cite{elecUK} (figure~1c). In both cases, the
competition between parties seems to play an important role, and the
assumption of non-interacting elements can be
unrealistic~\footnote{The effects of interaction are studied in
\cite{nota2}, where we go beyond the non-interacting system using a
microscopic description based on complex networks.}.

\subsection{Bulk and Zipf regimes}

The situation in which $N/\Omega\rightarrow \,{\rm constant}\, \ne 0$ as 
$N,\Omega\rightarrow\infty$ will be referred to herefrom as the \emph{bulk regime}.
Now, in a recent communication \cite{algo}, we show that Zipf's law 
($\gamma=1$ in~(\ref{eq1}) with a slope of $-1$ in the rank-plot) can be
derived from the extremization of Fisher's information with {\it no}
constraints. In the thermodynamic context studied here, the absence
of normalization can be understood as the inability of the system to
reach the thermodynamic limit, i.e. $N/\Omega\rightarrow0$ as
$N,\Omega\rightarrow\infty$. In this case the system can not follow
(\ref{cold}).  Zipf's law emerges as this behaviour of the
density in what we will accordingly denominate the \emph{Zipf
regime} ($N/\Omega\rightarrow0$). We digress on the conditions for
both regimes in the example discussed  below.

We have studied the system formed by all Physics journals~\cite{isi}
($N=310$) using their total number of cites as coordinate $k$. If a
journal receives more cites due to its popularity, it becomes even
more popular and therefore it will receive more cites. Under such
conditions proportional growth and scale invariance are expected.
Since we consider {\it all sub-fields} of Physics, correlation
effects are much lower than they would be should we only consider
journals pertaining to an specific sub-field. Accordingly,  the
non-interacting approximation seems to be realistic in this
instance. In figure~3 we depict the rank-plot of the number of
citations in Physic journals, and find a slope approaching $-1$ for
the most-cited journals in the logarithmic representation
(figure~3a)
and an slope in the vicinity of r $+1$ for the less-cited journals
(figure~3b).
For the central part of the distribution, the bulk density reaches a
value of $\rho_0\sim57$ (figure~3c).

\begin{figure}[t!]
\centering
\includegraphics[width=0.5\linewidth,clip=true]{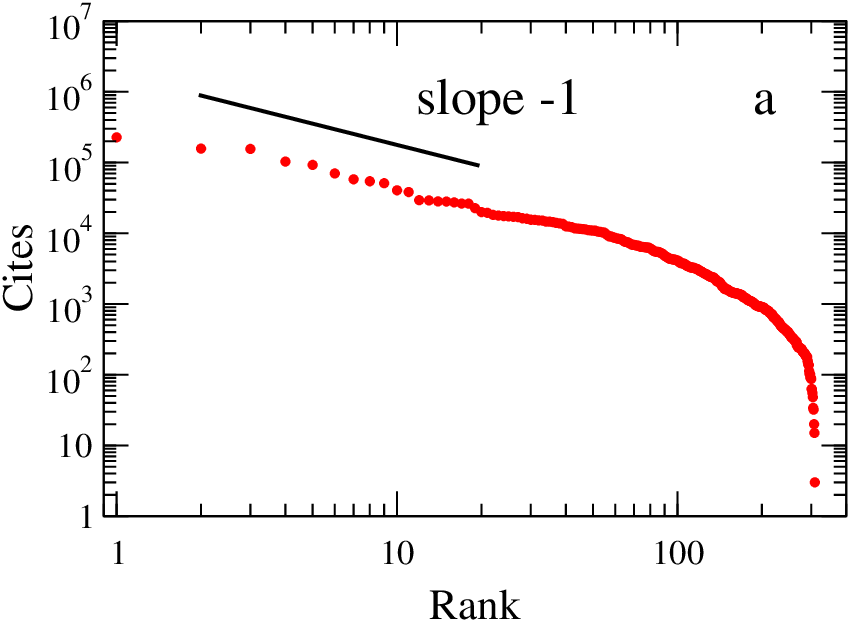}\\
\includegraphics[width=0.5\linewidth,clip=true]{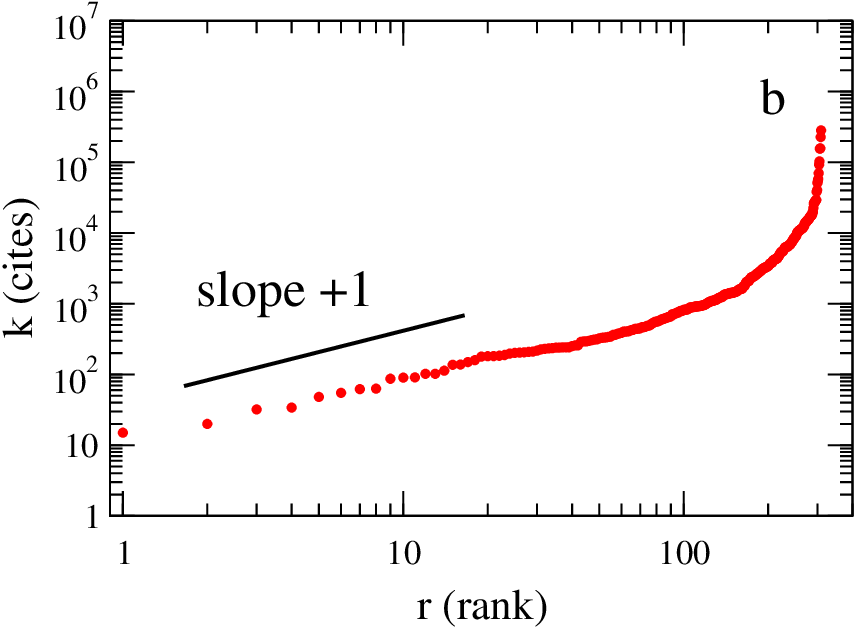}\\
\includegraphics[width=0.5\linewidth,clip=true]{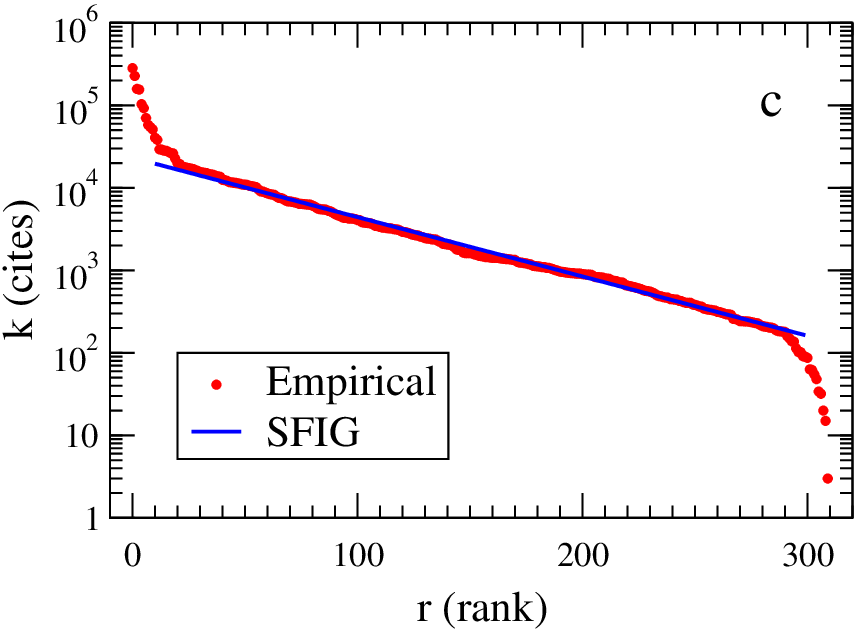}%
\caption{(colour on-line) \textbf{a}, rank-plot of the total number
of cites of physics journal, from most-cited to less-cited, in
logarithm scale. \textbf{b}, sorted from less-cited to most cited
\textbf{c}, same as a, in semi-logarithm scale. (Red dots: empirical
data; blue line: fit to (\ref{coldr})).\label{fig3}}
\end{figure}

This distribution shows a notably symmetric behaviour under
the change $k\rightarrow1/k$ ($u\rightarrow-u$). We exhibit in figure~4
the raw empirical data as compared with the distribution obtained from
the transformation $k'=c/k$ ($u'=-u+\ln c$), where
$c=3.3\times10^6$. The main part of the density distribution reaches
the bulk density obeying~(\ref{cold}), whereas Zipf's law emerges at
the edges, which could be understood as constituting the \emph{surface} of the
system, since they explain how the density (exponentially) falls from
its bulk-value to zero in $u$-space when the system is exposed to
an infinitely empty volume. This effect is clearly visible in
figure~5, where the empirical density distribution $p(u)du$ in 
$u$-space is compared with the ``fitted'' density
\begin{equation}\label{fit}
p(u)=\left\{
\begin{array}{ll}
\rho_Ze^{u-u_1}&\mathrm{if~} u<u_1\\
\rho_0&\mathrm{if~} u_1<u<u_2\\
\rho_Ze^{u_2-u}&\mathrm{if~} u>2
\end{array}
\right.
\end{equation}
where $\rho_Z=18$, $\rho_0=57$, $u_1=5.2$ and $u_2=10$.
These findings lead us to conclude that the system consisting of
Physics journals, when sorted by total number of citations, is a
perfect example of a finite scale-free ideal gas at equilibrium.

\begin{figure}[t!]
\center
\includegraphics[width=0.5\linewidth,clip=true]{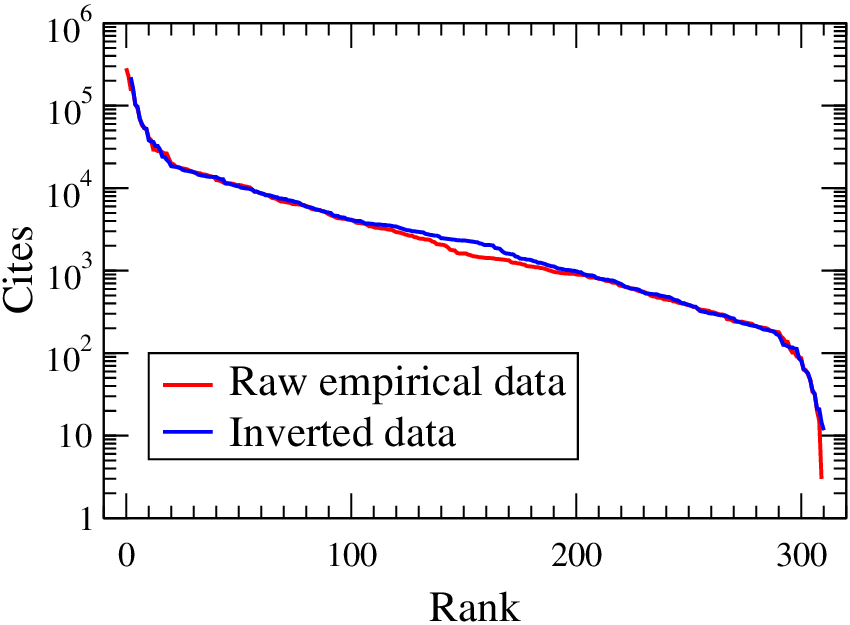}
\caption{(colour online) Rank-plot of the total number of cites of
Physics journal, from most-cited to less-cited, compared with the
distribution obtained from the inverse transformation
$k'=3.3\times10^6/k$ where $k$ is the number of cites.\label{fig4}}
\end{figure}

\begin{figure}[t!]
\center
\includegraphics[width=0.5\linewidth,clip=true]{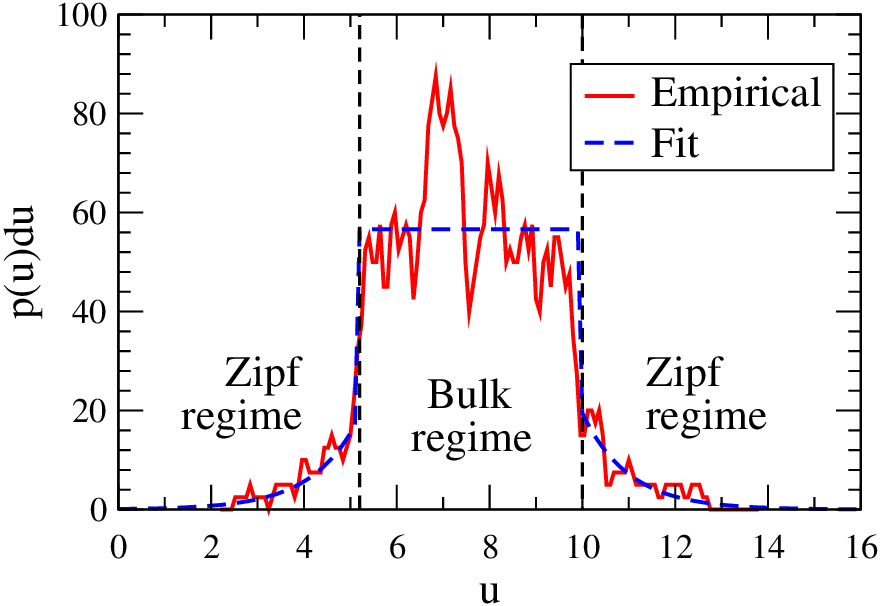}
\caption{(colour online) Empirical density distribution in $u$-space
of the total number of cites of Physics journals, compared with~(\ref{fit}). 
The bulk regime and the Zipf regime at the edges is
clearly visible. \label{fig5}}
\end{figure}

\subsection{An accompanying microscopic description} \label{p6}

The dynamics of the system under scrutiny here can be
microscopically described as a stochastic process using~(\ref{dyn})
together with the density distribution~(\ref{hw}). Treating $w$ as a
random variable, the pertinent stochastic equation of motion is
written in the guise of a geometrical Brownian motion, i.e.,~\cite{exp}
\begin{equation}\label{eqd}
dk = k\overline{w}dt + k\sigma_w dW,
\end{equation}
where $dW$ represents a Wiener process. In the $u$-space, this
equation reads
\begin{equation}\label{eqd2}
du = \overline{w}dt + \sigma_w dW,
\end{equation}
and is known to describe the celebrated Brownian motion~(\ref{eqd}),
which  exactly describes the dynamical condition found empirically
in~\cite{linux} and also the (stochastic) proportional growth model
used in~\cite{citis} to obtain Zipf's law. We thus dare to suggest
that we are dealing here with a sort of ``equivalent'' of a molecular
dynamics' simulation for gases/liquids~\cite{DM}.

Indeed, (\ref{eqd2}) describes $N$ Brownian walkers moving in a
fixed volume $\Omega$  with uniform density in $u$-space, a model
used in the literature to describe the ideal gas \cite{DM}. This
scenario can also be reproduced by our free-scale ideal gas merely
by choosing to represent the system with the coordinates $(k,v)$. In
figure~1a we show the rank-plot for $k$ of a system of $N=78$ geometrical
Brownian walkers with $\overline{w}=0.011$ and $\sigma_w=0.030$ in a volume $\Omega=4.5$, with 
d $k_1=200$ in reduced units, which approximately  describes the
distribution of the population of the province of Huelva. 
We also show in figure~2 the rankplot of $w$ of the same random walkers, 
compared with the empirical data.

\section{Conclusions}
\label{p7}

Our present considerations derive from the fact that, as shown in \cite{fisher2}, thermodynamics can be reformulated 
in terms of the minimization with appropriate constraints of Fisher's information. We have applied such reformulation in order to
 discuss the thermodynamics of scale-free systems and derived  the
density distribution in configuration space and the entropic expression
for the equilibrium  state of what we call SFIG: the scale-free ideal gas  (in the thermodynamic limit). We have encountered convincing 
 empirical evidences of the SFIG actual existence in sociological scenarios. Thus, we have dealt with city populations,
electoral results and citations in Physics journals. In such a  context it is seen that 
Zipf's law emerges naturally as the equilibrium density of the
non-interacting system when the volume grows without bounds, a situation that we call 
 the Zipf regime. Using empirical data we have revealed that this
regime can be understood as a density-decay at the ``surface''
separating the bulk from an empty and very large volume. Finally, we
have shown with a simulation of city-populations that  geometrical
Brownian motion can describe such systems at a microscopic level.

\section*{Acknowledgments}

We would like to thank D. Puigdomenech, D. Villuendas, M. Barranco,
R. Frieden, and B. H. Soffer for useful discussions. This work has
been partially performed under grant FIS2008-00421/FIS from DGI,
Spain (FEDER).

\end{document}